\documentstyle[letter]{ptptex}

\title{
Do relativistic corrections affect microlensing amplification?
}

\author{
Junya {\sc Ebina},$^{1}$ Takeaki {\sc Osuga},$^{1}$  
Hideki {\sc Asada}$^{1,2,}$\footnote{E-mail address: 
asada@phys.hirosaki-u.ac.jp} 
and Masumi {\sc Kasai}$^{1,}$\footnote{E-mail address: 
kasai@phys.hirosaki-u.ac.jp} 
}

\inst{%
$^1$ Faculty of Science and Technology, Hirosaki University, 
Hirosaki 036-8561, Japan\\
$^2$ 
Max-Planck-Institut f\"ur Astrophysik, 
Karl-Schwarzschild-Str. 1 \\
D-85741 Garching, Germany}

\abst{ Studies of gravitational microlensing are usually based on
  the lens equation, which is valid only to first order in the 
  gravitational constant $G$. However, the amplification factor of 
  microlensing is a second-order quantity with respect to
  $G$. Despite this fact, conventional studies are still based on the
  lowest-order lens equation. Then the question naturally arises: Why 
  are these conventional studies justified? We carefully study the 
  relativistic correction to the amplification factor at $O(G^2)$. 
  We show that the amplification factor for each image is corrected. 
  However, the total amplification remains unchanged at this order. 
}

\begin{document}

\maketitle

Microlensing has been one of most successful areas of study 
in astrophysics over the past decade, \cite{Paczynski96,NIK} 
since microlensing due to massive astrophysical compact halo 
objects (MACHO) were proposed by Paczynski \cite{Paczynski86} 
and detected first by MACHO \cite{Alcock} and EROS \cite{Aubourg} 
groups. 
The current microlensing searches are performed also 
in the direction of our galactic center.\cite{Udalski} 
Furthermore, there are ongoing/proposed projects for monitoring 
microlens events with short time intervals, say hours.\cite{Muraki} 
By such monitoring, we can quickly make follow-up observations 
with large telescopes, such as Keck, Subaru and the 
Very Large Telescope (VLT). 
Hence, we should be able to obtain (1) more accurate 
shapes of the light curve for microlens events and 
(2) a rapid increase in the number of events. 

In this paper, taking account of expected future precise 
measurements of the light curve, we study relativistic corrections 
to amplification by microlensing. 
Theoretical studies of the light curve are usually based on 
amplification calculated using the so-called deflection angle 
formula $4GM/bc^2$, where $M$ is the lens mass and $b$ is the impact 
parameter of a light ray. 
This deflection angle is calculated at linear order, $O(G)$, in the 
post-Minkowskian approximation, which is an expansion scheme 
in terms of $G$. 
Actually, only the Newtonian potential $GM/b$ is used. 
As for the amplification factor, however, it is necessary to calculate 
up to the second order $O(G^2)$, as shown in Eq. $(\ref{eq:amp})$ below. 
Despite this fact, most studies are still based on the lens equation, 
which is valid only to lowest order in $G$. 
In order to examine whether these conventional studies are
quantitatively correct, we study relativistic corrections to 
the amplification at the next order, $O(G^2)$. 
It is apparent that small changes could appear in the amplification. 
Rather surprisingly, however, we show that the total amplification 
remains unchanged at $O(G^2)$, while the amplifications for 
each image change. 
We use units in which $c=1$. 

First, we summarize the derivation of the total amplification 
for microlensing.\cite{Paczynski96} 
Denoting the source and image position angles by $\theta_S$ and 
$\theta_I$, respectively, the lens equation is written as 
\begin{equation}
\theta_S=\theta_I-\frac{D_{LS}}{D_S}\alpha , 
\end{equation}
where we have used the thin-lens approximation, and $D_{LS}$ 
and $D_S$ denote distances from the lens to the source 
and from the observer to the source, respectively. 
This gives us a mapping between the lens and source planes. 
We can consider the case $\theta_S \geq 0$ without loss of generality. 
For a point lens (generally a spherically symmetric lens), 
the deflection angle $\alpha$ becomes, at $O(G)$,  
\begin{equation}
\alpha=\frac{4GM}{b}. 
\end{equation}
Here we have assumed that the lens moves slowly, so that the effect 
due to its velocity is negligible. 
Then, the lens equation is
\begin{equation}
 \theta_S=\theta_I-\frac{\theta_E^2}{\theta_I}, 
\end{equation}
where $\theta_E$ is the angular radius of the  Einstein ring, 
\begin{equation}
\theta_E=\sqrt{\frac{4GMD_{LS}}{D_L D_S}},  
\label{eq:Einsteinring}
\end{equation}
and $D_L$ denotes the distance from the observer to the lens. 
For microlensing in our galaxy, the radius is typically 
\begin{equation}
\theta_E \sim 10^{-4} \Bigl( \frac{M}{0.1M_{\odot}} \Bigr)^{1/2} 
\Bigl( \frac{50\mbox{kpc}}{D_{S}} \Bigr)^{1/2} \mbox{arcsec.} , 
\end{equation}
where we have assumed $D_L \sim D_{LS}$. 

The lens equation is solved easily as 
\begin{equation}
\theta_{\pm}=\frac12(\theta_S \pm \sqrt{\theta_S^2+4 \theta_E^2}) . 
\end{equation}
{}From a simple geometrical argument, we find the amplification 
due to gravitational lensing to be 
\begin{equation}
A=\Bigl|\frac{\theta_I}{\theta_S}\frac{d\theta_I}{d\theta_S}\Bigr| 
 =  \frac{1}{\left|1 - \left(\frac{\theta_E}{\theta_I}\right)^4\right|}, 
\label{eq:amp}
\end{equation}
which is considered a function of the image position $\theta_I$. 
{}From Eqs.~(\ref{eq:Einsteinring}) and (\ref{eq:amp}), it is
apparent that the amplification factor $A$ is a second-order quantity
with respect to $G$. It should be noted that terms at first order 
in $G$ disappear in Eq. (7) as a consequence of cancellation. 
For each image $\theta_{\pm}$, the amplification factor becomes 
\begin{equation}
A_{\pm}=\frac{u^2+2}{2u\sqrt{u^2+4}} \pm \frac12 , 
\end{equation}
where $u$ denotes $\theta_S/\theta_E$, the source position 
in units of the Einstein ring radius. 
In typical microlensing events, the angular separation between 
the images is so small that all we can measure is the amplification 
of the total flux, 
\begin{equation}
A_{\mbox{total}}=A_{+} + A_{-}=\frac{u^2+2}{u\sqrt{u^2+4}}. 
\label{eq:amplification}
\end{equation}

Now, we are in the position to study the relativistic correction 
at $O(G^2)$. 
To this order, the correction to the deflection angle is \cite{ES} 
\begin{equation}
\Delta\alpha=\frac{15\pi G^2M^2}{4b^2} . 
\end{equation} 
Since the relativistic correction is generally small, 
it is convenient to treat the correction as a perturbation 
around the well-known results at $O(G)$. 
In order to do so, we define the angle in units of the 
Einstein ring radius as 
\begin{equation}
\tilde\theta=\frac{\theta}{\theta_E} . 
\end{equation}
Then, by noting that $\tilde\theta_I$ can be negative, 
we obtain the lens equation up to $O(G^2)$,  
\begin{equation}
\tilde\theta_S=\tilde\theta_I - \frac{1}{\tilde\theta_I} 
\Bigl( 1 + \frac{\lambda}{|\tilde\theta_I|} \Bigr) , 
\label{eq:lenseq}
\end{equation}
where we have defined the dimensionless parameter $\lambda$ as
\begin{equation}
\lambda=\frac{15\pi D_S \theta_E}{64 D_{LS}} . 
\end{equation}
This lens equation can be rewritten as 
\begin{equation}
\tilde\theta_I^3 - \tilde\theta_S \tilde\theta_I^2 
- \tilde\theta_I - \lambda \frac{\tilde\theta_I}{|\tilde\theta_I|} 
= 0 . 
\label{eq:lenseq2}
\end{equation}
Since $\theta_E$ is sufficiently small in most astronomical situations, 
we can take $\lambda$ as an expansion parameter.

The lens equation $(\ref{eq:lenseq2})$ is solved iteratively 
in terms of $\lambda$: 
For $\tilde\theta_I \geq 0$, the lens equation is rewritten as 
\begin{equation}
\tilde\theta_I^3 - \tilde\theta_S \tilde\theta_I^2 
- \tilde\theta_I - \lambda = 0 . 
\end{equation}
Substituting the form $\tilde\theta_{+} + \lambda\phi 
+ O(\lambda^2)$ for $\tilde\theta_I$ in this equation, 
we find the solution as 
\begin{equation}
\tilde\theta_{+}^{'} = \tilde\theta_{+} + 
\frac{\lambda}{\tilde\theta_{+}\sqrt{\tilde\theta_S^2 +4}} 
+ O(\lambda^2) , 
\end{equation}
where the prime denotes a quantity with relativistic corrections. 
For $\tilde\theta_I < 0$, the lens equation becomes 
\begin{equation}
\tilde\theta_I^3 - \tilde\theta_S \tilde\theta_I^2 
- \tilde\theta_I + \lambda = 0 . 
\end{equation}
which is solved up to $O(\lambda)$ as 
\begin{equation}
\tilde\theta_{-}^{'} = \tilde\theta_{-} +
\frac{\lambda}{\tilde\theta_{-}\sqrt{\tilde\theta_S^2 +4}} 
+ O(\lambda^2) . 
\end{equation}

The amplification for each image becomes 
\begin{equation}
A_{\pm}^{'} = \frac{u^2+2}{2u\sqrt{u^2+4}} \pm \Bigl( 
\frac12 - \lambda (u^2+4)^{-3/2} \Bigr) + O(\lambda^2) . 
\end{equation}
Hence, relativistic corrections at $O(G^2)$ change the amplification 
for each image. 
We find, however, the amplification for the total flux as 
\begin{equation}
A_{\mbox{total}}^{'} = \frac{u^2+2}{u\sqrt{u^2+4}} + O(\lambda^2) ,  
\end{equation}
which is the same as Eq. $(\ref{eq:amplification})$. 
Hence, in the microlensing, the total amplification remains 
unchanged, up to $O(G^2)$.  

For higher order relativistic corrections, terms of 
inverse powers in $\tilde\theta$ appear in Eq. $(\ref{eq:lenseq})$. 
In a similar manner to the present case of $O(G^2)$, we may find that 
for higher orders in relativistic corrections, only terms even  
in $\lambda$ in $A_{\mbox{total}}$ may appear, 
with every odd terms vanishing. 

Before closing this paper, we give a remark on $\theta_E$. 
The Einstein ring radius given by Eq. $(\ref{eq:Einsteinring})$ 
should be corrected at higher orders of $G$. 
Then, we may use the corrected radius $\theta_E^{'}$ 
in the analysis at $O(G^2)$. 
Only the ratio between the Einstein ring radius and the impact 
parameter determines the maximal amplification in light curves 
due to microlensing. 
However, since the impact parameter is not observable, a difference 
between $\theta_E$ and $\theta_E^{'}$ can be never observed. 
Therefore, the difference does not change our conclusion.

\section*{Acknowledgements}

We would like to thank M. Bartelmann for carefully reading the 
manuscript and helpful comments. 
H. A. would like to thank Gerhard B\"orner for hospitality at the 
Max-Planck-Institut f\"ur Astrophysik, where a part of this work 
was done. 
This work was supported in part by a Japanese Grant-in-Aid 
for Scientific Research from the Ministry of Education, Science, 
Sports and Culture, No. 11740130 (H. A.).

\end{document}